\documentclass[draft,grl]{agu2001}
\usepackage{graphicx}
\usepackage{amsmath}
\usepackage{amssymb}
\usepackage{xspace}
\usepackage[large]{subfigure}

\newcommand{\pdiff}[2][{}]{\ensuremath{\partial_{#2}\ifx\\#1\else^{#1}\fi}}

\newcommand{\unit}[2][{}]{\ensuremath{\ifx\\#1\else#1\,\fi\mathrm{#2}}}

\newcommand{\RS}{\mathrm{R_S}}
\newcommand{\RMP}{\mathrm{R_{MP}}}

\newcommand{\AMMX}{\textit{A10}\xspace}
\newcommand{\SMMX}{\textit{S10}\xspace}

\newcommand{\Cassini}{\textit{Cassini}\xspace}
\newcommand{\Voyager}{\textit{Voyager}\xspace}

\newcommand{\CAPS}{\renewcommand{\CAPS}{CAPS\xspace}\Cassini Plasma Spectrometer (CAPS)\xspace}
\newcommand{\MAG}{\renewcommand{\MAG}{MAG\xspace}\Cassini magnetometer (MAG)\xspace}
\newcommand{\MIMI}{\renewcommand{\MIMI}{MIMI\xspace}\Cassini Magnetospheric Imaging Instrument (MIMI)\xspace}
\newcommand{\INCA}{\renewcommand{\INCA}{INCA\xspace}ion-neutral camera (INCA)\xspace}
\newcommand{\PLS}{\renewcommand{\PLS}{PLS\xspace}\Voyager plasma science (PLS)\xspace}
\newcommand{\SLT}{\renewcommand{\SLT}{SLT\xspace}Saturn local time (SLT)\xspace}
\newcommand{\IBS}{\renewcommand{\IBS}{IBS\xspace}ion beam spectrometer (IBS)\xspace}
\newcommand{\IMS}{\renewcommand{\IMS}{IMS\xspace}ion mass spectrometer (IMS)\xspace}

\newcommand{\rev}[1]{\renewcommand{\rev}[1]{Rev~##1\xspace}Revolution~#1 (Rev~#1)\xspace}

\newcommand{\Fig}[1]{Fig.\,{#1}}

\newcommand{\revise}[1]{{#1}}
\newcommand{\emrevise}[1]{{#1}}

\newcommand{\comment}[1]{}

\authorrunninghead{ACHILLEOS ET AL.}

\titlerunninghead{SATURN'S HOT PLASMADISK}

\authoraddr{N. Achilleos, 
Atmospheric Physics Laboratory, University College London,
Gower Street, London WC1E 6BT, UK. (nick@apl.ucl.ac.uk)}

\begin{document}

%
%
%
%
%

%
%

\title{Influence of Hot Plasma Pressure on \emrevise{the} Global Structure of Saturn's Magnetodisk}
%

%
%




\authors{N. Achilleos, \altaffilmark{1}$^{^,}$\altaffilmark{3} 
P. Guio, \altaffilmark{1}$^{^,}$\altaffilmark{3} 
C. S. Arridge, \altaffilmark{2}$^{^,}$\altaffilmark{3}  
N. Sergis, \altaffilmark{4}
R.~J. Wilson, \altaffilmark{6}$^{^,}$\altaffilmark{5}
M.~F. Thomsen\altaffilmark{5}
and A.~J. Coates\altaffilmark{2}$^{^,}$\altaffilmark{3}  }

\altaffiltext{1} 
{Department of Physics and Astronomy, University College London, Gower Street, London, UK}

\altaffiltext{2} 
{Mullard Space Science Laboratory, Holmbury St. Mary, Dorking, Surrey, UK}

\altaffiltext{3} 
{Centre for Planetary Sciences at UCL / Birkbeck, University College London, Gower Street, London, UK}

\altaffiltext{4} 
{Office for Space Research and Technology, Academy of Athens, Athens, Greece}

\altaffiltext{5} 
{Space Science and Applications, Los Alamos National Laboratory, P.O. Box 1663, D466, Los Alamos, USA}

\altaffiltext{6} 
{Laboratory of Atmospheric and Space Physics, University of Colorado at Boulder, Boulder, USA}

%
%
%

%
%


\begin{abstract}
Using a model of force balance in Saturn's disk-like 
magnetosphere,
we show that variations in hot plasma pressure can change the magnetic field configuration.
This effect changes (i) the location of the
magnetopause, even at fixed solar wind dynamic pressure, 
and (ii) the magnetic mapping between ionosphere and disk.
The model uses equatorial observations as a boundary condition---we test its
predictions over a wide latitude range by comparison with 
a \Cassini high-inclination orbit of magnetic field and hot plasma pressure data.
We find reasonable agreement 
over time scales larger than the period of Saturn kilometric radiation (also known as the camshaft period). 
\end{abstract}

%
%

%

\begin{article}

%
%

\section{Introduction}

%
%


%
%
Saturn's equatorial, rotating plasma disk 
is threaded by a magnetic field which changes from a quasi-dipolar geometry, at distances
$\unit[{\lesssim}12]{R_S}$ (Saturn's radius \unit[]{R_S} = \unit[60280]{km}),
to a ``magnetodisk'' \comment{configuration }in the outer region,
where the radial field component is dominant at distances  
$\unit[{\sim}1\mbox{--}2]{R_S}$ from the equator (e.g\ \citet{arridge2007}). 
This magnetodisk is supported by an azimuthal ring current whose
solenoid-like field \revise{adds to} the planetary dipole.
In early studies, the ring current
was represented by a bound annular region with four free parameters: a uniform half-thickness,
inner and outer radii, and a current scaling parameter 
(e.g.\ \citet{connerney1981}). In this model, current density $J_{\phi}$ was assumed inversely
proportional to cylindrical radial distance $\rho$. This formalism (``CAN81 disk'') was recently adapted
to analyze \Cassini magnetic data, establishing how
ring current parameters depend on magnetopause size \citep{bunce2007}.

More recently, a formalism for calculating self-consistent field and plasma distributions for the 
Jovian magnetodisk
\citep{caudal1986} has been implemented for \revise{Saturn}
(\citet{achilleos2010}, hereafter \AMMX).
This approach 
\revise{integrates}\comment{uses equatorial plasma data as a boundary condition for integrating}
equations of force balance throughout a rotating, axisymmetric 
magnetosphere (\S\ref{sec:modinputs}). 
In \S\ref{sec:modoutputs}, we use the \AMMX model to show that the variability of hot plasma pressure,
as observed,
affects the magnetospheric field structure---increased hot plasma content generally producing a more 
``inflated'' or ``disk-like'' field geometry. 
To test the model's high-latitude structure, we compare its
\comment{ global magnetic and plasma pressures} \revise{outputs}
with field and particle data from the Rev 40 high-inclination orbit of \Cassini.
We discuss the agreement between model and data 
\comment{over sufficiently large time scales} 
and summarize
our conclusions in \S\ref{sec:summary}. 

\section{Magnetodisk Model Inputs}
\label{sec:modinputs}
\begin{figure}[h]
\vspace{-4cm}
\noindent\includegraphics[width=25pc]{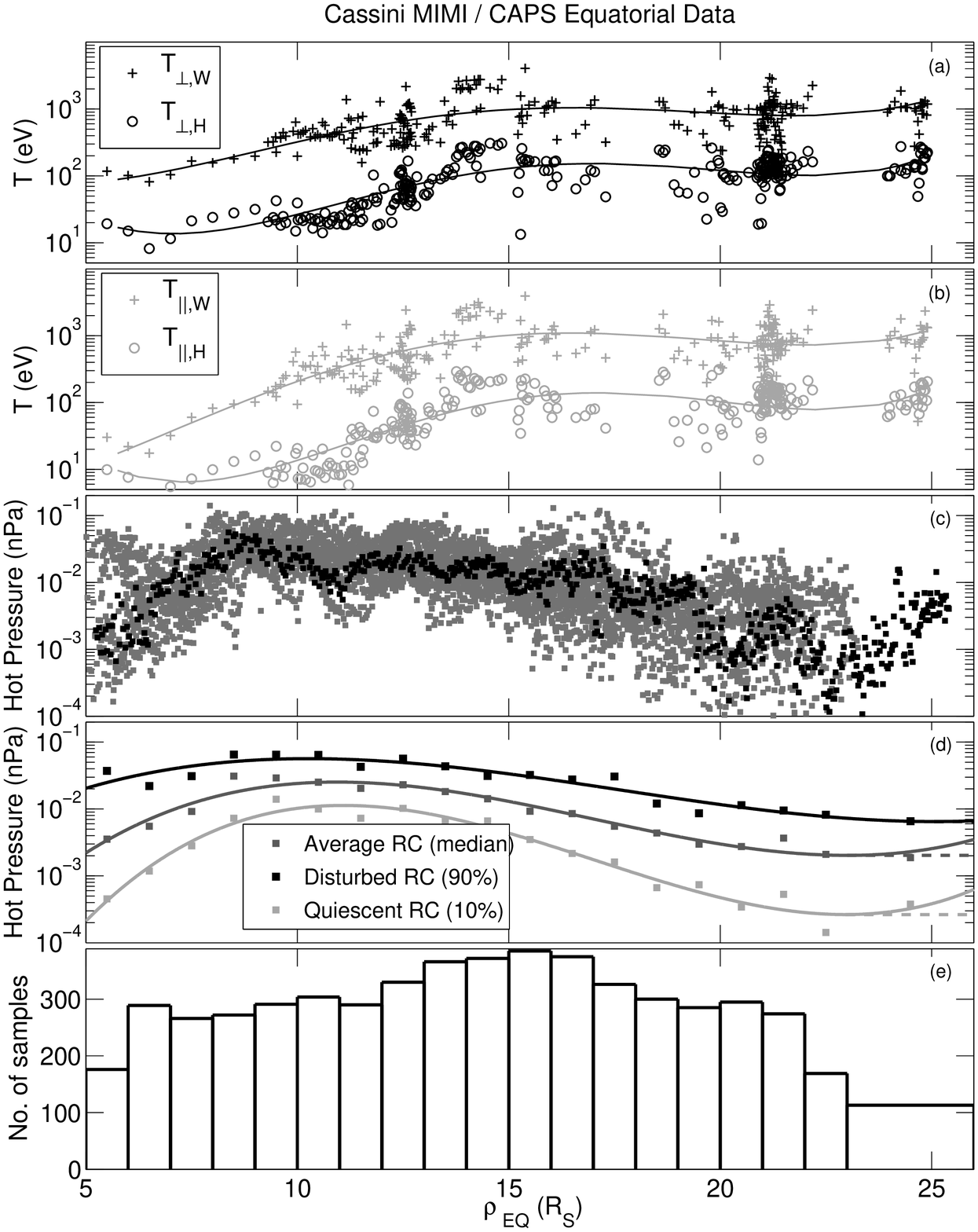}
\caption{\hspace{-0.5cm}(a), (b) Perpendicular and parallel temperature moments 
(with respect to assumed-southward magnetic field) 
for protons (H,circles) and water group ions
(W,crosses), from \citet{wilson2008} and \citet{mcandrews2009}.
Solid curves show quartic fits; (c) Hot plasma pressure derived 
from the \MIMI data. Gray squares cover 23 \Cassini orbits, black squares are from 
\comment{one of these (Rev 3)}\revise{Rev 3};
(d) Median and quantile pressures computed for bins in (e). Solid curves
are cubic fits used to calculate {$P_hV_{\alpha}$} (see text); (e) Radial
distance bins and number of samples used\comment{ to compute the parameters in (d)}.
}
\label{fig:plasmadata}
\end{figure}

We have improved \AMMX's representation of
the cold equatorial ion temperature over
\unit[10\mbox{--}25]{\RS}, by combining temperatures for protons
and water group ions from 
\citet{mcandrews2009} ($\unit[{\sim}10\mbox{--}30]{\RS}$) 
and \citet{wilson2008} (\unit[5\mbox{--}10]{\RS}),
who analyzed energy spectra acquired by the \IMS, a subsystem of the \CAPS \citep{young2004}.
Temperature moments versus equatorial distance, along with quartic fits,
are shown in \Fig{\ref{fig:plasmadata}a,b} (fit coefficients available on request). \AMMX's initial study
used only \citet{wilson2008}'s temperatures, with parabolic extrapolation
beyond \unit[10]{\RS}.
Using the updated profiles, we derived the single plasma temperature
$T_c(\rho)$ required by the model, whose field lines are loci of constant, isotropic temperature.
This was done as in \AMMX, by taking weighted average moments for each ion species,
then combining these according to a prescribed plasma composition.


To model hot plasma content, we used the 
equatorial product \mbox{$P_h(\rho)\,V_{\alpha}(\rho)$} 
of hot plasma pressure (assumed constant along a field line) and unit flux tube volume 
(\citet{caudal1986}, \AMMX). \comment{This captures realistic behavior of decreasing (increasing) 
interior plasma pressure for a flux tube which expands (contracts) during model computation.}We extended 
the \MIMI pressure dataset used in \AMMX's initial study 
\citep{sergis2007} to cover 23 near-equatorial orbits in total, 
between October, 2004 and March, 2006.
\revise{The hot pressure arises from
ions (mainly H+ and O+) with energies $\unit[{>}3]{keV}$ 
(e.g.\ \citet{sergis2009}).
The moments are
computed in the spacecraft frame, from an incomplete sampling of the phase space.
However, for $\rho{>}\unit[12]{\RS}$, energetic ions have
typical thermal speeds ${\sim}10$ times higher than the plasma flow speed.
The resulting hot pressures are uncertain by factors of ${\sim}2{-}3$,  
within the scatter of the presented data.}
We used pressure moments 
sampled every 10 minutes, \revise{taken}
\comment{along orbit segments }from the dayside, equatorial magnetosphere\comment{, defined by}: 
(i) absolute latitudes
${<}5^{\circ}$, (ii) Saturn local times 09:00--15:00, 
and (iii) locations between
each orbit's last inbound and \revise{next} outbound magnetopause crossings.

All hot pressures are shown as gray symbols
in \Fig{\ref{fig:plasmadata}c}, collapsed onto a single profile in
$\rho$. For comparison, data for
\Cassini Rev 3 (February 14--23, 2005) are overlain in black \revise{(data from this orbit
were previously analyzed by \AMMX)}. 
\Fig{\ref{fig:plasmadata}d} shows the median, 10\% and 90\% quantiles. 
\comment{for the samples 
within each of the distance bins shown in \Fig{\ref{fig:plasmadata}e}. }Solid curves 
are cubic fits to these parameters, 
which correspond to ``average'', ``disturbed'' and ``quiescent''
ring current states (following \citet{sergis2007} and \AMMX). We assumed a zero-slope profile
past $\unit[{\sim}23]{R_S}$ (dashed lines in \Fig{\ref{fig:plasmadata}d}), however this did not affect 
our conclusions. 
Repeating this parametrization\comment{ of hot pressure}, incorporating all local times, elevated the quantile pressures
by ${\sim}40\%$.

\section{Comparison of Magnetodisk Models and Observations}
\label{sec:modoutputs}
\begin{figure}[h]
\noindent\includegraphics[width=35pc]{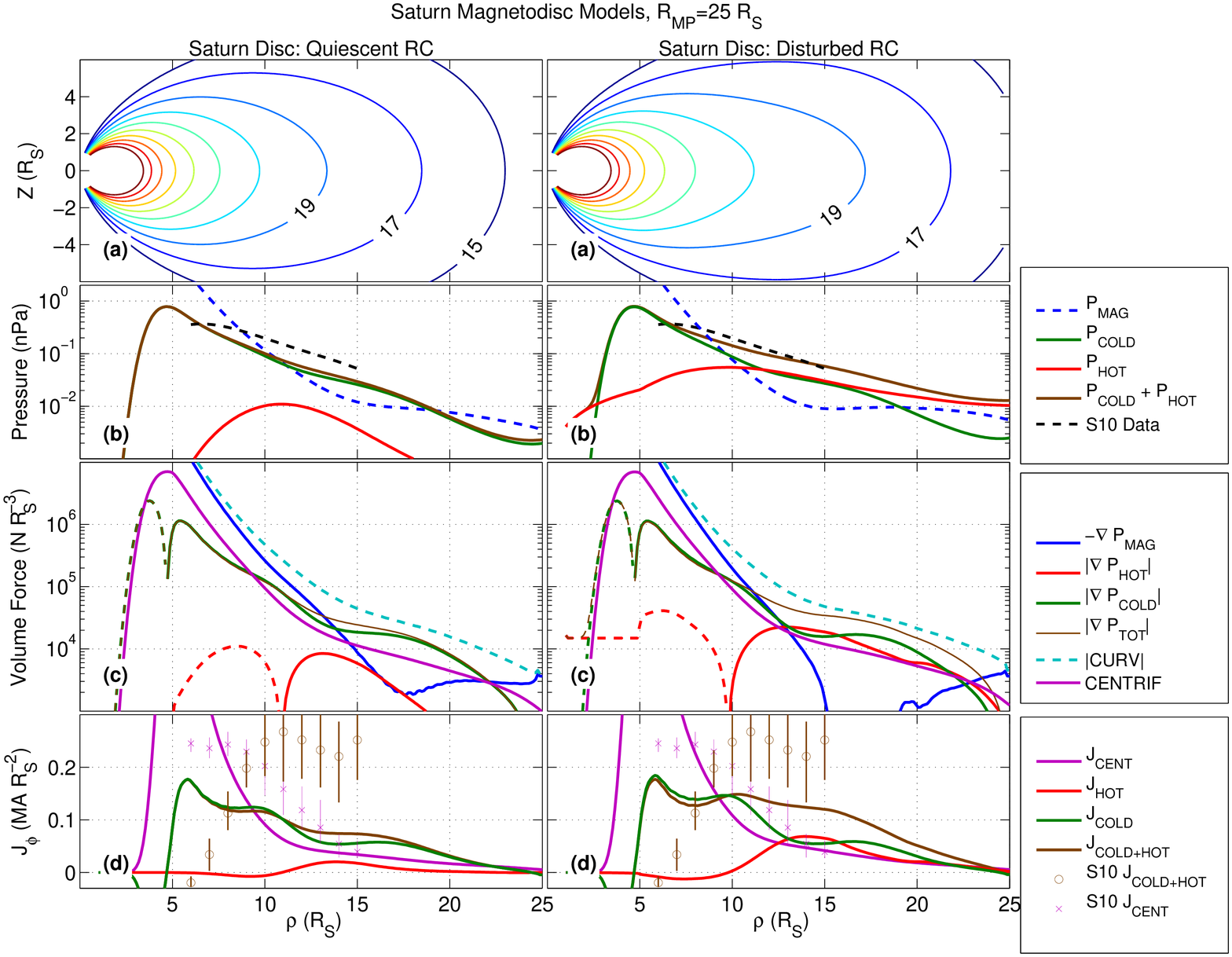}
\caption{Model outputs for \comment{average }magnetopause radius $\RMP=\unit[25]{\RS}$, and
for quiescent (left\comment{panels}) and disturbed (right\comment{panels}) ring current state.
(a) Magnetic field lines\comment{ (colored curves)}, partially labeled with the colatitude $\theta_i$ (in degrees) of 
their northern ionospheric footpoint \revise{($2^{\circ}$ steps in $\theta_i$)};
(b) Equatorial \comment{profiles of }pressure. Color code indicates \comment{either }magnetic or 
plasma pressure (cold, hot
and total),
dashed black curve is \revise{from}\comment{the profile of}  \SMMX;
(c) Equatorial volume forces, \revise{colored}\comment{color coded} according to physical origin. 
Line style indicates radially outward (solid) or inward (dashed) force;
(d) Equatorial \comment{profiles of }azimuthal current density, 
\revise{colored}\comment{color coded} according to physical 
origin, compared with the medians (symbols), 10\% and 90\% quantiles (vertical bars)
of the data-derived currents of \SMMX (\unit[1]{R_S} window).
}  
\label{fig:modeloutputs}
\end{figure}

\Fig{\ref{fig:modeloutputs}} illustrates the model magnetodisk's response to 
ring current state\comment{(i.e.\ hot plasma content)}. 
Left and right columns show calculations for fixed 
magnetopause radius $\RMP{=}\unit[25]{R_S}$ (average value for Saturn
\citep{achilleos2008}), but for quiescent and disturbed states (\Fig{\ref{fig:plasmadata}d}). 
\comment{The panels}\Fig{\ref{fig:modeloutputs}a}  
\revise{shows} that field lines
in the disturbed model become more \revise{radially} ``stretched'' for
${\rho}\unit[{\sim}10\mbox{--}15]{\RS}$. 
The labeled field lines intersect northern ionospheric colatitudes 
$\theta_i{=} 15^{\circ}$, $17^{\circ}$ and $19^{\circ}$.
\revise{These values are first order estimates, since we assume a centered dipole for
the planet's internal field; \citet{nichols2009} have used auroral observations and a more realistic
internal field model to show that the northern auroral oval has a radius ${\sim}1.5^{\circ}$
smaller than the southern. The model magnetopause shielding field is valid
for latitudes ${\lesssim}50^{\circ}$ \citep{caudal1986}, and
the labeled field lines exceed this limit for radial distances ${\lesssim}\unit[7]{R_S}$.}
Increased hot plasma content expands the equatorial distance of 
\revise{the labeled} field lines 
by $\unit[{\sim}3\mbox{--}4]{\RS}$. \revise{The ionospheric} colatitudes 
\revise{$15{-}17^{\circ}$} \revise{are ${\lesssim}1^{\circ}$ from}
the observed equatorward boundary
of the \revise{northern} auroral oval 
\revise{\citep{nichols2009}}. 
Thus
internal re-configuration
of hot plasma, at fixed
$\RMP$, can influence ionosphere-disk magnetic mapping 
and equatorial width of the 
auroral current layer. \revise{A similar effect was noted by \citet{grodent2008} from auroral
observations of Jupiter}.

In \Fig{\ref{fig:modeloutputs}b}, we plot equatorial magnetic and plasma pressure 
\comment{profiles}for both models. Field strength at $\rho{\gtrsim}\unit[15]{R_S}$
for the disturbed disk is more uniform than the quiescent model. Total model plasma pressure 
\revise{(disturbed)} agrees well, as shown, with that of \citet{sergis2010}
(hereafter \SMMX; derived directly from \CAPS and \MIMI pressures). The radial gradient of our
profile differs from that of \SMMX, however, which has consequences for azimuthal current (see below).
Our disturbed disk's maximum plasma $\beta$ of ${\sim}6.5$ at 
$\rho{\sim}\unit[14]{\RS}$ is
consistent with \SMMX's 
$\beta_{max}\unit[{\sim}\mbox{3--10}]{}$ for 
{$\rho{\sim}\unit[11\mbox{--}14]{\RS}$}.

\Fig{\ref{fig:modeloutputs}c} compares equatorial volume forces.
Again, the smaller gradient in the disturbed disk's magnetic pressure  
for $\rho{\gtrsim}\unit[15]{\RS}$ is evident. The corresponding increase in total plasma pressure gradient
is mainly balanced by increased curvature force of the more disk-like field. 
In the outer model disk,
plasma pressure gradient
becomes smaller than centrifugal force within a few \unit[]{R_S} of the
magnetopause. This result, arising from our improved pressure representation, 
differs greatly from \AMMX's ``transition distance'' of $\unit[{\sim}12]{\RS}$, but agrees
with \citet{arridge2007}, who deduced plasma pressure from
\Cassini magnetic data during current sheet encounters.

\Fig{\ref{fig:modeloutputs}d} compares the modeled equatorial
azimuthal current density, associated with pressure gradient and centrifugal force, with \SMMX's data-derived values. 
Considering centrifugal (inertial) current, the model is in reasonable agreement
with the data for $\rho{\gtrsim}\unit[6]{\RS}$. 
The discrepancy is mainly due to the difference between our approach (see \AMMX) and
\SMMX for computing currents.
For centrifugal current, we assume \AMMX's polynomial
fit to plasma angular velocity $\omega$ obtained from
high-energy particle anisotropies \citep{kane2008}, combined with inner magnetospheric
values from \citet{wilson2008}. Combining our \comment{theoretical }magnetic field model 
and profile for ion flux tube content
\revise{produces}\comment{allows} the centrifugal term\comment{ to be obtained}. 
\SMMX employed \CAPS density and velocity
moments from \citet{wilson2008} and \citet{mcandrews2009}, 
\revise{and}\comment{as well as} their own
\CAPS ion moment computations, in order to derive \revise{and fit} centrifugal volume force 
\comment{and fit this quantity}as a function of $\rho$. The ratio of our centrifugal \comment{volume }force
to that of \SMMX ranges from ${\sim}2$ at \unit[{\sim}6]{R_S} to ${\sim}0.5$ at \unit[{\sim}12]{R_S}.   

Considering pressure gradient, we recall
(discussion of \Fig{\ref{fig:modeloutputs}b})
considerable differences in plasma pressure gradients 
between our model and \SMMX's values.
Thus for $\rho{\gtrsim}\unit[8]{\RS}$, our disturbed model's 
total pressure current seems reasonably consistent with \SMMX's lower limits. 
Current density exceeds inertial at \unit[{\sim}8]{\RS} in both model 
and data.  For $\rho{\lesssim}\unit[8]{\RS}$, \citet{sergis2010} find ${>}25\%$ of total current 
arises from pressure anisotropy. This is not included in our model 
at present\comment{, hence we have scope to improve future comparisons}.

\begin{figure}[h]
\noindent\includegraphics[width=25pc]{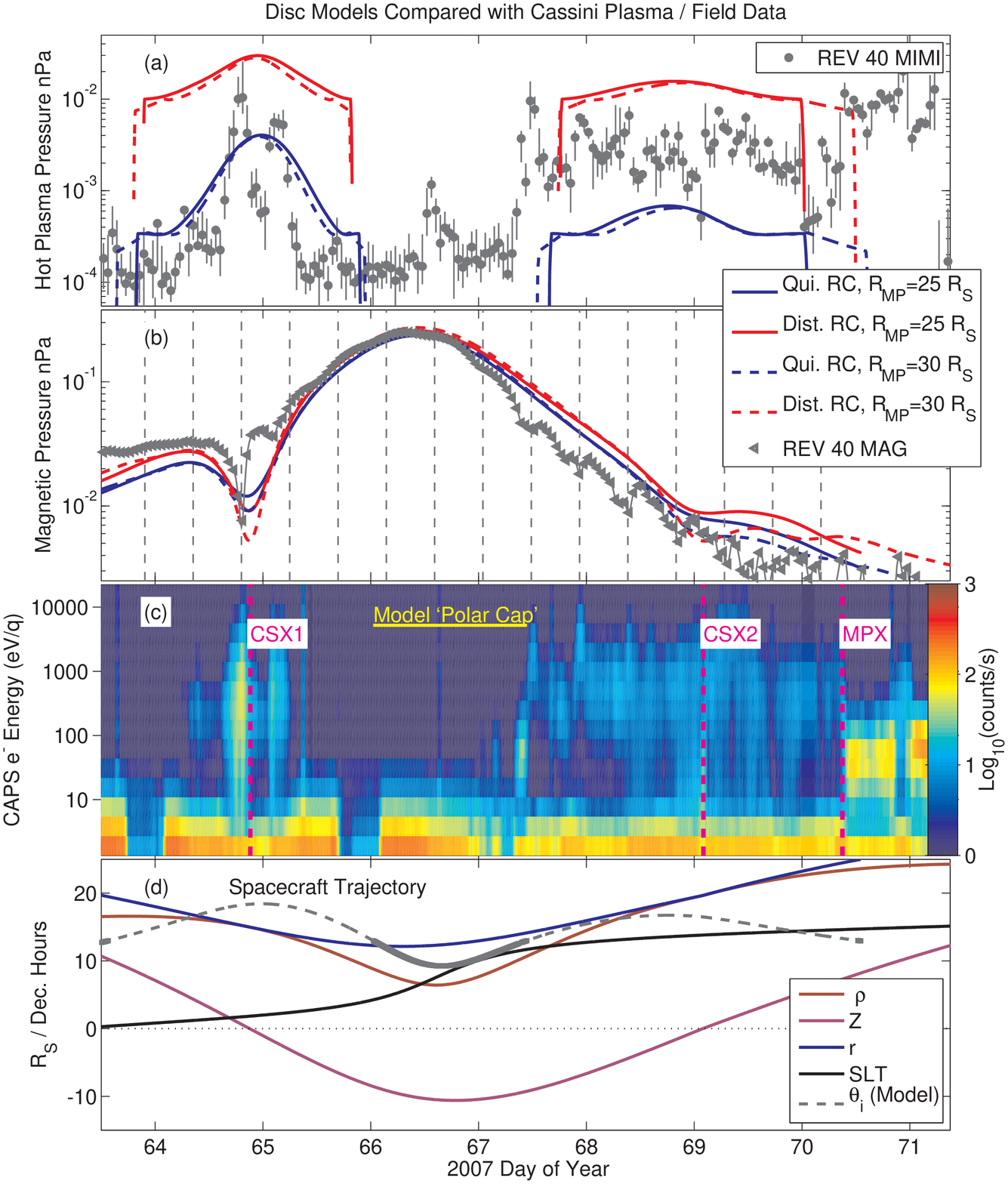}
\caption{(a) Hot plasma pressure versus time for \Cassini's Rev 40 orbit
(symbols are medians, vertical bars show 25\% and 75\% quantiles, for a 
one-hour window). Disk model pressures are overlaid (see code in panel (b)); 
(b) Observed and modeled magnetic pressure for Rev 40, with
\MAG data shown as symbols and model curves color- and line-coded according to magnetopause radius and
ring current activity. Thin vertical dashed lines are spaced by the planetary rotation period; 
(c) \CAPS electron spectrogram (one-hour average from Anode 5), with  
current sheet (CSX) and magnetopause (MPX) crossings marked. Horizontal line shows the model polar cap interval; 
(d) Spacecraft trajectory, showing cylindrical \revise{and spherical}
radial distance $\rho$ \revise{and $r$}, vertical coordinate $Z$, local time SLT and model colatitude 
$\theta_i$ of northern magnetic footpoint.}
\label{fig:modelrev40}
\end{figure}

In \Fig{\ref{fig:modelrev40}a,b}, we compare model outputs with hot plasma and magnetic
pressures for \Cassini's Rev 40 orbit from the \MIMI and \MAG \citep{dougherty2004}. The model assumes 
constant hot pressure 
$P_h$ along field lines, hence the
change in local pressure with altitude $Z$ (from the equator) is equivalent to the change in the value
$P_{h0}$ at the equatorial
crossings of the field lines sampled by the spacecraft. Values of $\RMP$ were determined by 
\AMMX, who modeled individual field components 
during this orbit\comment{(we consider only magnetic pressure herein)}.

Hot pressures near the nightside 
current sheet crossing (CSX1) are well captured by a quiescent disk model, while the dayside crossing (CSX2)
requires an ``intermediate'' state. Sharp ``dropouts'' in pressure near days ${\sim}67.5$ and 
${\sim}70$ (decimal day of year) are also seen in the model, and coincide with the spacecraft's exit from and
entry into the polar cap, characterized by open magnetic flux and relative absence of plasma. This region manifests in 
the model as field lines which do not cross the equator inside the magnetopause, 
hence are assumed to be devoid of plasma.
Polar cap boundary crossings are also seen in nightside models, near days ${\sim}63.9$ and ${\sim}65.9$.
For these local times, however, observed pressures are small, near noise level, and 
could obscure any dropout signature.
Quasi-periodic ``pulsations'' are present in plasma and magnetic data, which we emphasize with
vertical lines spaced by the nominal \unit[10.75]{hr} planetary rotation period.
These features are due to the quasi-periodic ``camshaft'' oscillations in the Kronian
magnetosphere, whose period agrees with that of the Saturn kilometric radiation 
(e.g.\ \citet{djsmgk2007,kurth2007,provan2009}).

The change in magnetic pressure when crossing the current sheet is greater at 
CSX1 than CSX2. \comment{The model fields 
(\Fig{\ref{fig:modeloutputs}}a) indicate the reason.}CSX1 occurs at $\rho{\sim}\unit[15]{R_S}$
in the ``middle magnetosphere'' where the field lines are most disk-like, hence produce 
a strong contrast in field strength between the inside and outside of the sheet 
\revise{(\Fig{\ref{fig:modeloutputs}}a)}.
CSX2 occurs at $\rho{\sim}\unit[20]{R_S}$
towards the ``outer region'' where field lines are more ``blunt'' (larger radius of curvature),
this contrast is thus reduced, and the cold plasma sheet is thicker (\AMMX). 
Models with more disturbed disk \mbox{and\,/\,or} 
larger $\RMP$ show a change in magnetic pressure comparable to that
seen at CSX1. However, the observed sheet transit is more rapid
than the model analog, \revise{perhaps}\comment{This could be } due to 
plasma sheet motion linked to camshaft oscillations. 

\Fig{\ref{fig:modelrev40}c} is a \CAPS electron spectrogram for the same orbit. 
The \comment{observed }decrease in cold electron plasma (counts) with distance from the
equator is more pronounced near CSX1 than CSX2. This aspect reflects
global force balance, which allows the cold population to maintain higher pressure for a given $Z$ in regions
where the field lines have larger curvature radii, and centrifugal confinement is weaker (e.g.\ \AMMX, \citet{caudal1986}).
Nightside data are better matched by a more quiescent disk and larger $\RMP$, both possible
signatures of a thinner nightside plasma sheet, as described by \citet{sergis2009}. 
\comment{Further modeling of this
aspect is thus warranted.
The main polar cap interval again shows a comparatively empty region.}
The magnetosheath population seen after the outbound magnetopause
crossing (MPX, early on day 70) shows peak energies similar to those near the exit from the polar cap (via the
dayside cusp) at day ${\sim}67.5$.
Trajectory information in \Fig{\ref{fig:modelrev40}d} includes 
northern ionospheric
colatitude $\theta_i$ magnetically conjugate to \Cassini's location. Regions with
$\theta_i{<}13^{\circ}$ are shown as a solid curve
(also shown in \Fig{\ref{fig:modelrev40}c}), and represent a model polar cap,
located on field lines
poleward of the auroral oval boundary
\citep{badman2006}.

\section{Conclusions}
\label{sec:summary}
Saturn's magnetospheric field can be significantly
modified by internal changes in hot plasma pressure. 
\comment{A useful, related study would be
orbit-by-orbit correlation analysis of disk field strength and hot plasma content, which would
determine useful proxies for these quantities.}Magnetic mapping between ionosphere and
disk is also dependent on internal
plasma state (as well as external solar wind pressure, e.g \citet{bunce2008}). This result suggests a corresponding
influence on the distribution of currents which flow between these regions and their associated auroral emissions.
Comparing our models with both field and plasma data allows identification of ring current state and 
\revise{its magnetic signatures.}
\comment{magnetopause radius, as well as 
different magnetospheric plasma regimes.}


%
%

%
%

\begin{acknowledgments}
We acknowledge the continued collaboration of the \Cassini magnetometer (MAG) and
plasma (CAPS, MIMI) instrument teams. CSA was supported by an STFC Postdoctoral Fellowship 
under grant number ST/G007462/1
\end{acknowledgments}

\end{article}

\end{document}